\documentclass[twocolumn,prl,aps,showpacs,floatfix]{revtex4}
\usepackage{amsmath}
\usepackage{graphicx}

\begin{document}
\title{

{
\centerline{\normalsize\hfill \tt SFB/CPP-06-24}
\centerline{\normalsize\hfill \tt TTP06-17}
\centerline{\normalsize\hfill \tt hep-ph/0605201}
\baselineskip 11pt
}

\vspace{1cm}
Four-Loop QCD Corrections to the $\boldsymbol{\rho}$ Parameter
}
\author{K.G.~Chetyrkin${}^{1*}$}
%\thanks{{\small Permanent address:
%Institute for Nuclear Research, Russian Academy of Sciences,
% Moscow 117312, Russia}}
\author{M.~Faisst${}^1$}
\author{J.H.~K\"uhn${}^1$}
\author{P.~Maierh\"ofer${}^1$}
\author{C.~Sturm${}^{2}$}
\affiliation{${}^1$Institut f\"ur Theoretische Teilchenphysik,
  Universit\"at Karlsruhe, D-76128 Karlsruhe, Germany}
\affiliation{${}^{2}$Dipartimento di Fisica Teorica, Universit{\`a} di
  Torino, Italy \& INFN, Sezione di Torino, Italy}

\begin{abstract}
\noindent
The four-loop QCD corrections to the electroweak $\rho$ parameter 
arising from top and bottom quark loops are computed. Specifically we
evaluate the missing ``non-singlet'' piece. Using algebraic methods the
amplitude is reduced to a set of  around  50 new   master integrals
which are calculated with various analytical and numerical
methods. Inclusion  of the newly completed term halves the final value
of the four-loop   correction for the minimally renormalized top-quark
mass.  The predictions for the shift of the weak mixing angle  and  the
W-boson mass is thus stabilized.

\end{abstract}


\maketitle

Electroweak precision measurements and calculations provide stringent
and decisive tests of the quantum fluctuations  predicted from quantum
field theory. As a most notable example, the indirect determination of
the top quark mass, $m_t$, mainly through its contribution to the $\rho$
parameter \cite{unknown:2005di}, coincides remarkably well with the mass
measurement performed by the CDF and D0 experiments at the TEVATRON
\cite{Group:2006qt}. Along the same line, the bounds on the mass of the
Higgs-boson depend critically on the knowledge of $m_t$ and the control
of the top-mass dependent effects on precision observables.

A large group of   dominant radiative corrections can be absorbed in the
shift of the $\rho$ parameter from its lowest order value
$\rho_{\text{Born}}=1$. The result for the one-loop approximation 
\begin{equation}
\delta \rho = 3 \, x_t = 3 \, \frac{G_F m_t^2}{8\sqrt{2}\, \pi^2},
\end{equation}
hence  quadratic in $m_t$, was first evaluated in \cite{Veltman:1977kh}
and used to establish a limit on the mass splitting within one fermion
doublet. In order to make full use of the present experimental
precision, this one-loop calculation has been improved by two-loop
\cite{Djouadi:1987gn,Djouadi:1987di,Kniehl:1988ie}  and even three-loop
QCD  corrections \cite{Avdeev:1994db,Chetyrkin:1995ix}. Also important
are two-loop \cite{vanderBij:1986hy,Barbieri:1992nz, Barbieri:1992dq,%
Fleischer:1993ub, Fleischer:1994cb} and three-loop
\cite{vanderBij:2000cg,Faisst:2003px}   electroweak effects proportional
to $x_t^2$ and $x_t^3$, respectively, and the three-loop mixed
corrections of order $\alpha_s x_t^2$ \cite{Faisst:2003px}.

An important ingredient for the interpretation of these results in terms
of top mass measurements performed at hadron colliders or at a future
linear collider is the relation between the pole mass and the
$\overline{\mathrm{MS}}$-mass definitions, the former being useful for
the determination of $m_t$ at colliders, the latter being employed in
actual calculations and in short-distance considerations. To match the
present three-loop  precision of the $\rho$ parameter, this relation
must be know in two-loop approximation
\cite{Gray:1990yh,Jegerlehner:2003sp,Faisst:2004gn,Eiras:2005yt}, 
and for the four-loop calculation under discussion the corresponding
three-loop result
\cite{Chetyrkin:1999ys,Chetyrkin:1999qi,Melnikov:2000qh} must be
employed.

For fixed pole mass of the top quark, the three-loop result leads to a
shift of about 10 MeV in the mass of the $W$-boson as discussed in
\cite{Chetyrkin:1995js,Faisst:2003px}. 
(This applies both to the pure QCD corrections and the mixed
QCD-electroweak one.)  Conversely, the corresponding shift of the top
quark pole mass amounts to 1.5 GeV. (Similar considerations apply to
the effective weak mixing angle and other precision observables.)
These values are comparable to the experimental precision anticipated
for top- and $W$-mass measurements at the International Linear
Collider \cite{Aguilar-Saavedra:2001rg}.  
In addition, there exists a
disagreement (on the level of $3\sigma$) between the values of the 
so-called on-shell week mixing angle, $\sin^2 \theta_W$,  as measured
by NuTeV collaboration in deep-ienalstic neutrino scattering  \cite{Zeller:2001hh}
and as obtained from the global fit \cite{Rosenbleck:2005fn} of the Standard Model 
to the electroweak precision   data \footnote{For more detailed discussion of 
various aspects of the issue see e.g. works 
\cite{Kataev:2002sj,Kulagin:2003wz,McFarland:2003jw,Diener:2005me,Erler:2006vt} 
and references therein.}. 

From all these considerations an improvement of the theoretical accuracy seems, 
therefore, desirable.
This, however, requires the evaluation of four-loop QCD
corrections to the $\rho$ parameter, the topic of the present work.

The shift in the  $\rho$ parameter  is given by
\begin{equation}
  \label{eq:rhodef}
  \delta \rho = \frac{\Pi_T^Z(0)}{M_Z^2} -
  \frac{\Pi_T^W(0)}{M_W^2},
\end{equation}
where $\Pi^{W/Z}_T(0)$ are the transversal parts of the $W$- and 
$Z$-boson self-energies at zero momentum transfer, respectively.
The calculation is thus reduced to the evaluation of vacuum (tadpole)
diagrams.

The  $W$-self-energy receives contributions from the correlator of the
``non-diagonal'' $t$-$b$-current only. Contributions to the
$Z$ self-energy originate only from the ``diagonal'' axial current
correlator induced by top quark loops. The vector current part
vanishes due to current conservation, the bottom quark is taken as
massless. The nonvanishing parts of $\Pi_T^Z(0)$ are conveniently
decomposed into non-singlet and singlet pieces characterized by
Feynman diagrams where the external current couples to the same and to
two different closed fermion lines, respectively. In three-loop
approximation the singlet-piece is larger than the non-singlet piece
by nearly a factor twenty. This has motivated the authors of
\cite{Schroder:2005db} to evaluate, in a first step, the four-loop
singlet piece. The strategy employed in that paper was based on the
algebraic reduction of the amplitudes to a small set of
master-integrals with the help of the Laporta algorithm
\cite{Laporta:1996mq,Laporta:2001dd} an approach which was recently
also applied to the evaluation of the two lowest non-vanishing Taylor
coefficients of the vacuum polarization
\cite{Chetyrkin:2006xg,Boughezal:2006px} and to the decoupling of
heavy quarks in QCD
\cite{Chetyrkin:2006pe,Schroder:2005hy,Chetyrkin:2005ia},
both in four-loop approximation.

In order to complete the evaluation of the four-loop QCD corrections to
the $\rho$ parameter, $\Pi^{W}_T(0)$ and the non-singlet parts of
$\Pi^{Z}_T(0)$ are required. 
The evaluation of $\Pi^{Z}_T(0)$ is fairly straightforward:
the input diagrams have been generated with QGRAF \cite{Nogueira:1991ex}; 
for the algebraic reduction to master integrals an efficient
program has been constructed \cite{Sturm:diss} which relies 
on  FORM3 and FERMAT \cite{Vermaseren:2000nd,Lewis,Tentyukov:2006ys}.
Furthermore, the full set of the corresponding  master integrals 
is available with high precision \cite{Schroder:2005va,Chetyrkin:2006dh}.

The evaluation
of $\Pi^{W}_T(0)$, however, requires the knowledge of a sizeable number
of new master-integrals, a major part  of  them (around 40) nontrivial
to evaluate precisely. The master integrals can be chosen in many
different ways. As discussed in  \cite{Chetyrkin:2006dh} the choice of a
so called ``$\epsilon$-finite basis'' leads to integrals particularly
suited for the evaluation through   Pad\'e approximations. On the other
hand, topologies with eight lines or less are conveniently calculated
through difference equations. In the present paper we therefore employ a
combined approach, which makes use of difference equations
\cite{Laporta:2001dd,Laporta:2002pg}  to evaluate the
simpler topologies, i.\ e. those with up to eight lines, and a
semi-numerical method  based on  Pad\'e approximations
\cite{Fleischer:1994ef,Baikov:1995ui,ChFK:LL04,Faisst:diss,Chetyrkin:2006dh}. There,
a suitably chosen line of the four-loop vacuum diagram is cut, the
large- and the small $q^2$ behaviour of the resulting three-loop
propagator are calculated analytically
\cite{Gorishnii:1989gt,Steinhauser:2000ry,Harlander:1997zb,%
Seidensticker:1999bb}, the function in the whole region is represented
by Pad\'e approximations and the remaining $q^2$ integration is
performed numerically (see Fig. 1).  

\begin{widetext}

\begin{center}
\begin{minipage}[b]{14cm}
  \begin{center}
    \includegraphics{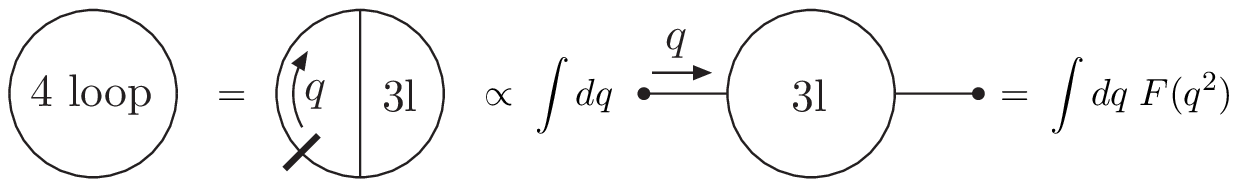}
     \end{center}
\end{minipage}
\end{center}

Fig 1. Symbolic description of the Pad\'e method: One line of the
vacuum integral is cut, the resulting propagator is represented by a 
 Pad\'e approximation and integrated numerically.

\end{widetext}

An estimate of the numerical uncertainty is obtained by
comparing different Pad\'e approximations based on the same input
information from the large and small $q^2$-region, or by increasing the
input information through inclusion of more terms from the high and the
low $q^2$ region. Furthermore, in all  cases at least two
different lines were cut to check the consistency of the results.
A detailed discussion of the various applications,
characteristic examples and comparisons with analytic results, e.g. for
the lowest moment of the polarization function can be
found in \cite{Faisst:diss}.

In contrast to the applications discussed in earlier publications
\cite{ChFK:LL04,Faisst:diss,Chetyrkin:2006dh}, massless cuts unavoidably
arise in some  of the relevant diagrams and a generalization of the
method is required: In addition to the introduction of a suitably chosen
function needed for the subtraction of the high energy logarithms, another
function is employed for the subtraction of the logarithms arising 
from the massless cut in the low energy limit.

The result for the shift in the $\rho$ parameter can be cast into the
following form:
\begin{equation}
  \label{eq:rho-define}
\delta \rho^{\overline{\mathrm{MS}}} = 
 3 x_t \sum_{i=0}^{3} \left(\frac{\alpha_s}{\pi}\right)^i
 \delta \rho^{\overline{\mathrm{MS}}}_i
\end{equation}

Here $x_t$ is expressed in terms of the $\overline{\mathrm{MS}}$ quark
mass $m_t\equiv m_t(\mu)$ at scale $\mu=m_t$, and $\alpha_s$, defined in
the $\overline{\mathrm{MS}}$-scheme for six flavors, is chosen at the
same scale. The normalization factors are such that $\delta
\rho^{\overline{\mathrm{MS}}}_0=1$.

For the four-loop non-singlet result, decomposed according to the
various color structures and the $n_f$-dependence, we find:
\begin{eqnarray}
&{}&\delta \rho^{\overline{\mathrm{MS}}}_3(\mbox{non-singlet}) = 
\\
&{}&
\nonumber
1.5211  \, C_F^3
+1.2363\, C_F^2\, C_A 
\\
&{}&
\nonumber
 -2.3132 \, C_F^2 \,T\, n_l 
-4.5962 \, C_F^2 \,T \,n_h 
\\
&{}&
\nonumber
 +0.7438\, C_F\, C_A^2 
-1.3705\, C_F\, C_A \,T\, n_l 
\\
&{}&
\nonumber
+2.5037\, C_F\, C_A \,T \, n_h 
+0.4681\, C_F \, T^2 \, n_l^2 
\\
&{}&
\nonumber
+0.6880\, C_F \, T^2 \, n_h^2 
 +\, 0.8495 \, C_F \, T^2 \, n_h n_l  
{},
\label{eq:rho-color}
\end{eqnarray}
with $C_F=(N_c^2-1)/(2N_c)$, $C_A=N_c$, and $T=1/2$, where $N_c=3$ is
the number of colors. $n_f$ denotes the number of active (light plus
heavy) quark fields, with $n_f= n_l + n_h$.  This result has been also
independently obtained with the  help of  direct  application of the
Pa\'de approximation   method like  it was described in
\cite{ChFK:LL04,Faisst:diss} for  the lowest Taylor coefficients of the
vacuum polarization. We find agreement  for all color structures with
the  relative accuracy  varying between  0.4\% and 4\%, and confirm the
result for the singlet contribution.

Setting $n_h=1$, $n_l=5$ and the color coefficients to their natural
values, we find 
\begin{eqnarray}
  \label{eq:rho-number}
\delta \rho^{\overline{\mathrm{MS}}}_3      & =
\nonumber
\delta \rho^{\overline{\mathrm{MS}}}_3(\mbox{singlet})
 +\delta \rho^{\overline{\mathrm{MS}}}_3(\mbox{non-singlet})\\
&    = -3.2866 + 1.6067     = -1.6799
{},
\end{eqnarray}
where we have  also displayed the result of \cite{Schroder:2005db} for the 
singlet piece. The singlet piece is still  larger  than the non-singlet
piece by a factor two. Nevertheless, the hierarchy is less pronounced
than in the three-loop case. Numerically, the overall correction looks
%tiny
 small, just as in the two- and three-loop case. However, if the result is
expressed in terms of the pole mass, a major shift originates from the
large correction in the pole-$\overline{\mathrm{MS}}$ relation:  
\begin{equation}
  \label{eq:rho-pole}
\delta \rho^{pole}_3 = - 93.1501
{}.
\end{equation}
For fixed top mass, this corresponds to a shift of around 2  MeV  in the $W$-boson
mass, well below the precision anticipated for future experiments.

In conclusion,  the full ${\cal O}(X_t \alpha_s^3)$ contribution  to  the  $\rho$ 
parameter proves to be  small and  the result based on the three-loop calculation is stabilized.

{\em Acknowledgments:} We would like to thank P. Baikov and V. Smirnov
for useful cross checks of some diagrams and M. Steinhauser for support
in the usage of MATAD. This work has been supported by DFG through
SFB/TR~9 and by BMBF, Grant No. 05HT4VKA/3.
The work of C.S. was also partially supported by MIUR
under contract 2001023713$\_$006.
\vspace{1cm}

\noindent
{\bf Note added.}
\\
\noindent
The results of our calculations have been recently confirmed in the independent 
work \cite{Boughezal:2006xk}.
\vspace{1cm}

{${}^*${\small Permanent address:
Institute for Nuclear Research, Russian Academy of Sciences,
 Moscow 117312, Russia.}}


\begin{thebibliography}{56}
\expandafter\ifx\csname natexlab\endcsname\relax\def\natexlab#1{#1}\fi
\expandafter\ifx\csname bibnamefont\endcsname\relax
  \def\bibnamefont#1{#1}\fi
\expandafter\ifx\csname bibfnamefont\endcsname\relax
  \def\bibfnamefont#1{#1}\fi
\expandafter\ifx\csname citenamefont\endcsname\relax
  \def\citenamefont#1{#1}\fi
\expandafter\ifx\csname url\endcsname\relax
  \def\url#1{\texttt{#1}}\fi
\expandafter\ifx\csname urlprefix\endcsname\relax\def\urlprefix{URL }\fi
\providecommand{\bibinfo}[2]{#2}
\providecommand{\eprint}[2][]{\url{#2}}

\bibitem[{\citenamefont{[ LEP Electroweak Working
      Group]}(2005)}]{unknown:2005di}
  \bibinfo{author}{\bibfnamefont{}\bibnamefont{LEP Electroweak Working
      Group}}  \eprint{hep-ex/0511027}.

%  \bibitem[{unk(2005)}]{unknown:2005di}
%  (\bibinfo{year}{2005}), \eprint{hep-ex/0511027}.



\bibitem[{\citenamefont{[ Tevatron Electroweak Working Group
      ]}(2006)}]{Group:2006qt}
  \bibinfo{author}{\bibfnamefont{}\bibnamefont{Tevatron Electroweak
      Working Group}} (\bibinfo{year}{2006}),
  \eprint{hep-ex/0603039}.


%\bibitem[{\citenamefont{Group}(2006)}]{Group:2006qt}
%\bibinfo{author}{\bibfnamefont{T.~E.~W.} \bibnamefont{Group}}
%  (\bibinfo{year}{2006}), \eprint{hep-ex/0603039}.


\bibitem[{\citenamefont{Veltman}(1977)}]{Veltman:1977kh}
\bibinfo{author}{\bibfnamefont{M.~J.~G.} \bibnamefont{Veltman}},
  \bibinfo{journal}{Nucl. Phys.} \textbf{\bibinfo{volume}{B123}},
  \bibinfo{pages}{89} (\bibinfo{year}{1977}).

\bibitem[{\citenamefont{Djouadi and Verzegnassi}(1987)}]{Djouadi:1987gn}
\bibinfo{author}{\bibfnamefont{A.}~\bibnamefont{Djouadi}} \bibnamefont{and}
  \bibinfo{author}{\bibfnamefont{C.}~\bibnamefont{Verzegnassi}},
  \bibinfo{journal}{Phys. Lett.} \textbf{\bibinfo{volume}{B195}},
  \bibinfo{pages}{265} (\bibinfo{year}{1987}).

\bibitem[{\citenamefont{Djouadi}(1988)}]{Djouadi:1987di}
\bibinfo{author}{\bibfnamefont{A.}~\bibnamefont{Djouadi}},
  \bibinfo{journal}{Nuovo Cim.} \textbf{\bibinfo{volume}{A100}},
  \bibinfo{pages}{357} (\bibinfo{year}{1988}).

\bibitem[{\citenamefont{Kniehl et~al.}(1988)\citenamefont{Kniehl, K{\"u}hn, and
  Stuart}}]{Kniehl:1988ie}
\bibinfo{author}{\bibfnamefont{B.~A.} \bibnamefont{Kniehl}},
  \bibinfo{author}{\bibfnamefont{J.~H.} \bibnamefont{K{\"u}hn}},
  \bibnamefont{and} \bibinfo{author}{\bibfnamefont{R.~G.}
  \bibnamefont{Stuart}}, \bibinfo{journal}{Phys. Lett.}
  \textbf{\bibinfo{volume}{B214}}, \bibinfo{pages}{621} (\bibinfo{year}{1988}).

\bibitem[{\citenamefont{Avdeev et~al.}(1994)\citenamefont{Avdeev, Fleischer,
  Mikhailov, and Tarasov}}]{Avdeev:1994db}
\bibinfo{author}{\bibfnamefont{L.}~\bibnamefont{Avdeev}},
  \bibinfo{author}{\bibfnamefont{J.}~\bibnamefont{Fleischer}},
  \bibinfo{author}{\bibfnamefont{S.}~\bibnamefont{Mikhailov}},
  \bibnamefont{and} \bibinfo{author}{\bibfnamefont{O.}~\bibnamefont{Tarasov}},
  \bibinfo{journal}{Phys. Lett.} \textbf{\bibinfo{volume}{B336}},
  \bibinfo{pages}{560} (\bibinfo{year}{1994}), \eprint{hep-ph/9406363}.

\bibitem[{\citenamefont{Chetyrkin
  et~al.}(1995{\natexlab{a}})\citenamefont{Chetyrkin, K{{\"u}}hn, and
  Steinhauser}}]{Chetyrkin:1995ix}
\bibinfo{author}{\bibfnamefont{K.~G.} \bibnamefont{Chetyrkin}},
  \bibinfo{author}{\bibfnamefont{J.~H.} \bibnamefont{K{{\"u}}hn}},
  \bibnamefont{and}
  \bibinfo{author}{\bibfnamefont{M.}~\bibnamefont{Steinhauser}},
  \bibinfo{journal}{Phys. Lett.} \textbf{\bibinfo{volume}{B351}},
  \bibinfo{pages}{331} (\bibinfo{year}{1995}{\natexlab{a}}),
  \eprint{hep-ph/9502291}.

\bibitem[{\citenamefont{van~der Bij and Hoogeveen}(1987)}]{vanderBij:1986hy}
\bibinfo{author}{\bibfnamefont{J.~J.} \bibnamefont{van~der Bij}}
  \bibnamefont{and}
  \bibinfo{author}{\bibfnamefont{F.}~\bibnamefont{Hoogeveen}},
  \bibinfo{journal}{Nucl. Phys.} \textbf{\bibinfo{volume}{B283}},
  \bibinfo{pages}{477} (\bibinfo{year}{1987}).

\bibitem[{\citenamefont{Barbieri et~al.}(1992)\citenamefont{Barbieri, Beccaria,
  Ciafaloni, Curci, and Vicere}}]{Barbieri:1992nz}
\bibinfo{author}{\bibfnamefont{R.}~\bibnamefont{Barbieri}},
  \bibinfo{author}{\bibfnamefont{M.}~\bibnamefont{Beccaria}},
  \bibinfo{author}{\bibfnamefont{P.}~\bibnamefont{Ciafaloni}},
  \bibinfo{author}{\bibfnamefont{G.}~\bibnamefont{Curci}}, \bibnamefont{and}
  \bibinfo{author}{\bibfnamefont{A.}~\bibnamefont{Vicere}},
  \bibinfo{journal}{Phys. Lett.} \textbf{\bibinfo{volume}{B288}},
  \bibinfo{pages}{95} (\bibinfo{year}{1992}), \eprint{hep-ph/9205238}.

\bibitem[{\citenamefont{Barbieri et~al.}(1993)\citenamefont{Barbieri, Beccaria,
  Ciafaloni, Curci, and Vicere}}]{Barbieri:1992dq}
\bibinfo{author}{\bibfnamefont{R.}~\bibnamefont{Barbieri}},
  \bibinfo{author}{\bibfnamefont{M.}~\bibnamefont{Beccaria}},
  \bibinfo{author}{\bibfnamefont{P.}~\bibnamefont{Ciafaloni}},
  \bibinfo{author}{\bibfnamefont{G.}~\bibnamefont{Curci}}, \bibnamefont{and}
  \bibinfo{author}{\bibfnamefont{A.}~\bibnamefont{Vicere}},
  \bibinfo{journal}{Nucl. Phys.} \textbf{\bibinfo{volume}{B409}},
  \bibinfo{pages}{105} (\bibinfo{year}{1993}).

\bibitem[{\citenamefont{Fleischer et~al.}(1993)\citenamefont{Fleischer,
  Tarasov, and Jegerlehner}}]{Fleischer:1993ub}
\bibinfo{author}{\bibfnamefont{J.}~\bibnamefont{Fleischer}},
  \bibinfo{author}{\bibfnamefont{O.~V.} \bibnamefont{Tarasov}},
  \bibnamefont{and}
  \bibinfo{author}{\bibfnamefont{F.}~\bibnamefont{Jegerlehner}},
  \bibinfo{journal}{Phys. Lett.} \textbf{\bibinfo{volume}{B319}},
  \bibinfo{pages}{249} (\bibinfo{year}{1993}).

\bibitem[{\citenamefont{Fleischer et~al.}(1995)\citenamefont{Fleischer,
  Tarasov, and Jegerlehner}}]{Fleischer:1994cb}
\bibinfo{author}{\bibfnamefont{J.}~\bibnamefont{Fleischer}},
  \bibinfo{author}{\bibfnamefont{O.~V.} \bibnamefont{Tarasov}},
  \bibnamefont{and}
  \bibinfo{author}{\bibfnamefont{F.}~\bibnamefont{Jegerlehner}},
  \bibinfo{journal}{Phys. Rev.} \textbf{\bibinfo{volume}{D51}},
  \bibinfo{pages}{3820} (\bibinfo{year}{1995}).

\bibitem[{\citenamefont{van~der Bij et~al.}(2001)\citenamefont{van~der Bij,
  Chetyrkin, Faisst, Jikia, and Seidensticker}}]{vanderBij:2000cg}
\bibinfo{author}{\bibfnamefont{J.~J.} \bibnamefont{van~der Bij}},
  \bibinfo{author}{\bibfnamefont{K.~G.} \bibnamefont{Chetyrkin}},
  \bibinfo{author}{\bibfnamefont{M.}~\bibnamefont{Faisst}},
  \bibinfo{author}{\bibfnamefont{G.}~\bibnamefont{Jikia}}, \bibnamefont{and}
  \bibinfo{author}{\bibfnamefont{T.}~\bibnamefont{Seidensticker}},
  \bibinfo{journal}{Phys. Lett.} \textbf{\bibinfo{volume}{B498}},
  \bibinfo{pages}{156} (\bibinfo{year}{2001}), \eprint{hep-ph/0011373}.

\bibitem[{\citenamefont{Faisst et~al.}(2003)\citenamefont{Faisst, K{\"u}hn,
  Seidensticker, and Veretin}}]{Faisst:2003px}
\bibinfo{author}{\bibfnamefont{M.}~\bibnamefont{Faisst}},
  \bibinfo{author}{\bibfnamefont{J.~H.} \bibnamefont{K{\"u}hn}},
  \bibinfo{author}{\bibfnamefont{T.}~\bibnamefont{Seidensticker}},
  \bibnamefont{and} \bibinfo{author}{\bibfnamefont{O.}~\bibnamefont{Veretin}},
  \bibinfo{journal}{Nucl. Phys.} \textbf{\bibinfo{volume}{B665}},
  \bibinfo{pages}{649} (\bibinfo{year}{2003}), \eprint{hep-ph/0302275}.

\bibitem[{\citenamefont{Gray et~al.}(1990)\citenamefont{Gray, Broadhurst,
  Grafe, and Schilcher}}]{Gray:1990yh}
\bibinfo{author}{\bibfnamefont{N.}~\bibnamefont{Gray}},
  \bibinfo{author}{\bibfnamefont{D.~J.} \bibnamefont{Broadhurst}},
  \bibinfo{author}{\bibfnamefont{W.}~\bibnamefont{Grafe}}, \bibnamefont{and}
  \bibinfo{author}{\bibfnamefont{K.}~\bibnamefont{Schilcher}},
  \bibinfo{journal}{Z. Phys.} \textbf{\bibinfo{volume}{C48}},
  \bibinfo{pages}{673} (\bibinfo{year}{1990}).

\bibitem[{\citenamefont{Jegerlehner and Kalmykov}(2003)}]{Jegerlehner:2003sp}
\bibinfo{author}{\bibfnamefont{F.}~\bibnamefont{Jegerlehner}} \bibnamefont{and}
  \bibinfo{author}{\bibfnamefont{M.~Y.} \bibnamefont{Kalmykov}},
  \bibinfo{journal}{Acta Phys. Polon.} \textbf{\bibinfo{volume}{B34}},
  \bibinfo{pages}{5335} (\bibinfo{year}{2003}), \eprint{hep-ph/0310361}.

\bibitem[{\citenamefont{Faisst et~al.}(2004{\natexlab{a}})\citenamefont{Faisst,
  K{\"u}hn, and Veretin}}]{Faisst:2004gn}
\bibinfo{author}{\bibfnamefont{M.}~\bibnamefont{Faisst}},
  \bibinfo{author}{\bibfnamefont{J.~H.} \bibnamefont{K{\"u}hn}},
  \bibnamefont{and} \bibinfo{author}{\bibfnamefont{O.}~\bibnamefont{Veretin}},
  \bibinfo{journal}{Phys. Lett.} \textbf{\bibinfo{volume}{B589}},
  \bibinfo{pages}{35} (\bibinfo{year}{2004}{\natexlab{a}}),
  \eprint{hep-ph/0403026}.

\bibitem[{\citenamefont{Eiras and Steinhauser}(2006)}]{Eiras:2005yt}
\bibinfo{author}{\bibfnamefont{D.}~\bibnamefont{Eiras}} \bibnamefont{and}
  \bibinfo{author}{\bibfnamefont{M.}~\bibnamefont{Steinhauser}},
  \bibinfo{journal}{JHEP} \textbf{\bibinfo{volume}{02}}, \bibinfo{pages}{010}
  (\bibinfo{year}{2006}), \eprint{hep-ph/0512099}.

\bibitem[{\citenamefont{Chetyrkin and Steinhauser}(1999)}]{Chetyrkin:1999ys}
\bibinfo{author}{\bibfnamefont{K.~G.} \bibnamefont{Chetyrkin}}
  \bibnamefont{and}
  \bibinfo{author}{\bibfnamefont{M.}~\bibnamefont{Steinhauser}},
  \bibinfo{journal}{Phys. Rev. Lett.} \textbf{\bibinfo{volume}{83}},
  \bibinfo{pages}{4001} (\bibinfo{year}{1999}), \eprint{hep-ph/9907509}.

\bibitem[{\citenamefont{Chetyrkin and Steinhauser}(2000)}]{Chetyrkin:1999qi}
\bibinfo{author}{\bibfnamefont{K.~G.} \bibnamefont{Chetyrkin}}
  \bibnamefont{and}
  \bibinfo{author}{\bibfnamefont{M.}~\bibnamefont{Steinhauser}},
  \bibinfo{journal}{Nucl. Phys.} \textbf{\bibinfo{volume}{B573}},
  \bibinfo{pages}{617} (\bibinfo{year}{2000}), \eprint{hep-ph/9911434}.

\bibitem[{\citenamefont{Melnikov and Ritbergen}(2000)}]{Melnikov:2000qh}
\bibinfo{author}{\bibfnamefont{K.}~\bibnamefont{Melnikov}} \bibnamefont{and}
  \bibinfo{author}{\bibfnamefont{T.~v.} \bibnamefont{Ritbergen}},
  \bibinfo{journal}{Phys. Lett.} \textbf{\bibinfo{volume}{B482}},
  \bibinfo{pages}{99} (\bibinfo{year}{2000}), \eprint{hep-ph/9912391}.

\bibitem[{\citenamefont{Chetyrkin
  et~al.}(1995{\natexlab{b}})\citenamefont{Chetyrkin, K{{\"u}}hn, and
  Steinhauser}}]{Chetyrkin:1995js}
\bibinfo{author}{\bibfnamefont{K.~G.} \bibnamefont{Chetyrkin}},
  \bibinfo{author}{\bibfnamefont{J.~H.} \bibnamefont{K{{\"u}}hn}},
  \bibnamefont{and}
  \bibinfo{author}{\bibfnamefont{M.}~\bibnamefont{Steinhauser}},
  \bibinfo{journal}{Phys. Rev. Lett.} \textbf{\bibinfo{volume}{75}},
  \bibinfo{pages}{3394} (\bibinfo{year}{1995}{\natexlab{b}}),
  \eprint{hep-ph/9504413}.

\bibitem[{\citenamefont{Aguilar-Saavedra
  et~al.}(2001)}]{Aguilar-Saavedra:2001rg}
\bibinfo{author}{\bibfnamefont{J.~A.} \bibnamefont{Aguilar-Saavedra}}
  \bibnamefont{et~al.} (\bibinfo{collaboration}{ECFA/DESY LC Physics Working
  Group}) (\bibinfo{year}{2001}), \eprint{hep-ph/0106315}.

\bibitem[{\citenamefont{Zeller et~al.}(2002)}]{Zeller:2001hh}
\bibinfo{author}{\bibfnamefont{G.~P.} \bibnamefont{Zeller}}
  \bibnamefont{et~al.} (\bibinfo{collaboration}{NuTeV}),
  \bibinfo{journal}{Phys. Rev. Lett.} \textbf{\bibinfo{volume}{88}},
  \bibinfo{pages}{091802} (\bibinfo{year}{2002}), \eprint{hep-ex/0110059}.

\bibitem[{\citenamefont{Rosenbleck}(2005)}]{Rosenbleck:2005fn}
\bibinfo{author}{\bibfnamefont{C.}~\bibnamefont{Rosenbleck}}
  (\bibinfo{year}{2005}), \eprint{hep-ex/0505033}.

\bibitem[{\citenamefont{Schr{\"o}der and Steinhauser}(2005)}]{Schroder:2005db}
\bibinfo{author}{\bibfnamefont{Y.}~\bibnamefont{Schr{\"o}der}}
  \bibnamefont{and}
  \bibinfo{author}{\bibfnamefont{M.}~\bibnamefont{Steinhauser}},
  \bibinfo{journal}{Phys. Lett.} \textbf{\bibinfo{volume}{B622}},
  \bibinfo{pages}{124} (\bibinfo{year}{2005}), \eprint{hep-ph/0504055}.

\bibitem[{\citenamefont{Laporta and Remiddi}(1996)}]{Laporta:1996mq}
\bibinfo{author}{\bibfnamefont{S.}~\bibnamefont{Laporta}} \bibnamefont{and}
  \bibinfo{author}{\bibfnamefont{E.}~\bibnamefont{Remiddi}},
  \bibinfo{journal}{Phys. Lett.} \textbf{\bibinfo{volume}{B379}},
  \bibinfo{pages}{283} (\bibinfo{year}{1996}), \eprint{hep-ph/9602417}.

\bibitem[{\citenamefont{Laporta}(2000)}]{Laporta:2001dd}
\bibinfo{author}{\bibfnamefont{S.}~\bibnamefont{Laporta}},
  \bibinfo{journal}{Int. J. Mod. Phys.} \textbf{\bibinfo{volume}{A15}},
  \bibinfo{pages}{5087} (\bibinfo{year}{2000}), \eprint{hep-ph/0102033}.

\bibitem[{\citenamefont{Chetyrkin
  et~al.}(2006{\natexlab{a}})\citenamefont{Chetyrkin, K{\"u}hn, and
  Sturm}}]{Chetyrkin:2006xg}
\bibinfo{author}{\bibfnamefont{K.~G.} \bibnamefont{Chetyrkin}},
  \bibinfo{author}{\bibfnamefont{J.~H.} \bibnamefont{K{\"u}hn}},
  \bibnamefont{and} \bibinfo{author}{\bibfnamefont{C.}~\bibnamefont{Sturm}}
  (\bibinfo{year}{2006}{\natexlab{a}}), \eprint{hep-ph/0604234}.

\bibitem[{\citenamefont{Boughezal et~al.}(2006)\citenamefont{Boughezal, Czakon,
  and Schutzmeier}}]{Boughezal:2006px}
\bibinfo{author}{\bibfnamefont{R.}~\bibnamefont{Boughezal}},
  \bibinfo{author}{\bibfnamefont{M.}~\bibnamefont{Czakon}}, \bibnamefont{and}
  \bibinfo{author}{\bibfnamefont{T.}~\bibnamefont{Schutzmeier}}
  (\bibinfo{year}{2006}), \eprint{hep-ph/0605023}.

\bibitem[{\citenamefont{Chetyrkin
  et~al.}(2006{\natexlab{b}})\citenamefont{Chetyrkin, K{\"u}hn, and
  Sturm}}]{Chetyrkin:2006pe}
\bibinfo{author}{\bibfnamefont{K.~G.} \bibnamefont{Chetyrkin}},
  \bibinfo{author}{\bibfnamefont{J.~H.} \bibnamefont{K{\"u}hn}},
  \bibnamefont{and} \bibinfo{author}{\bibfnamefont{C.}~\bibnamefont{Sturm}}
  (\bibinfo{year}{2006}{\natexlab{b}}), \eprint{hep-ph/0604187}.

\bibitem[{\citenamefont{Schr{\"o}der and Steinhauser}(2006)}]{Schroder:2005hy}
\bibinfo{author}{\bibfnamefont{Y.}~\bibnamefont{Schr{\"o}der}}
  \bibnamefont{and}
  \bibinfo{author}{\bibfnamefont{M.}~\bibnamefont{Steinhauser}},
  \bibinfo{journal}{JHEP} \textbf{\bibinfo{volume}{01}}, \bibinfo{pages}{051}
  (\bibinfo{year}{2006}), \eprint{hep-ph/0512058}.

\bibitem[{\citenamefont{Chetyrkin
  et~al.}(2006{\natexlab{c}})\citenamefont{Chetyrkin, K{\"u}hn, and
  Sturm}}]{Chetyrkin:2005ia}
\bibinfo{author}{\bibfnamefont{K.~G.} \bibnamefont{Chetyrkin}},
  \bibinfo{author}{\bibfnamefont{J.~H.} \bibnamefont{K{\"u}hn}},
  \bibnamefont{and} \bibinfo{author}{\bibfnamefont{C.}~\bibnamefont{Sturm}},
  \bibinfo{journal}{Nucl. Phys.} \textbf{\bibinfo{volume}{B744}},
  \bibinfo{pages}{121} (\bibinfo{year}{2006}{\natexlab{c}}),
  \eprint{hep-ph/0512060}.

\bibitem[{\citenamefont{Nogueira}(1993)}]{Nogueira:1991ex}
\bibinfo{author}{\bibfnamefont{P.}~\bibnamefont{Nogueira}},
  \bibinfo{journal}{J. Comput. Phys.} \textbf{\bibinfo{volume}{105}},
  \bibinfo{pages}{279} (\bibinfo{year}{1993}).

\bibitem[{\citenamefont{Sturm}(2005)}]{Sturm:diss}
\bibinfo{author}{\bibfnamefont{C.}~\bibnamefont{Sturm}}, \bibinfo{journal}{PhD
  thesis, Cuvillier Verlag, Goettingen} \textbf{\bibinfo{volume}{ISBN
  3-86537-569-X}} (\bibinfo{year}{2005}).

\bibitem[{\citenamefont{Vermaseren}(2000)}]{Vermaseren:2000nd}
\bibinfo{author}{\bibfnamefont{J.~A.~M.} \bibnamefont{Vermaseren}}
  (\bibinfo{year}{2000}), \eprint{math-ph/0010025}.


\bibitem[{\citenamefont{Lewis}()}]{Lewis}
\bibinfo{author}\mbox{{\mbox{\bibfnamefont{R.~H.} \bibnamefont{Lewis}}},
  \bibinfo{journal} \mbox{Fermat's User Guide,}} \\
  http://www.bway.net/\symbol{126}lewis/.


\bibitem[{\citenamefont{Tentyukov and Vermaseren}(2006)}]{Tentyukov:2006ys}
\bibinfo{author}{\bibfnamefont{M.}~\bibnamefont{Tentyukov}} \bibnamefont{and}
  \bibinfo{author}{\bibfnamefont{J.~A.~M.} \bibnamefont{Vermaseren}}
  (\bibinfo{year}{2006}), \eprint{cs.sc/0604052}.

\bibitem[{\citenamefont{Schr{\"o}der and Vuorinen}(2005)}]{Schroder:2005va}
\bibinfo{author}{\bibfnamefont{Y.}~\bibnamefont{Schr{\"o}der}}
  \bibnamefont{and} \bibinfo{author}{\bibfnamefont{A.}~\bibnamefont{Vuorinen}}
  (\bibinfo{year}{2005}), \eprint{hep-ph/0503209}.

\bibitem[{\citenamefont{Chetyrkin
  et~al.}(2006{\natexlab{d}})\citenamefont{Chetyrkin, Faisst, Sturm, and
  Tentyukov}}]{Chetyrkin:2006dh}
\bibinfo{author}{\bibfnamefont{K.~G.} \bibnamefont{Chetyrkin}},
  \bibinfo{author}{\bibfnamefont{M.}~\bibnamefont{Faisst}},
  \bibinfo{author}{\bibfnamefont{C.}~\bibnamefont{Sturm}}, \bibnamefont{and}
  \bibinfo{author}{\bibfnamefont{M.}~\bibnamefont{Tentyukov}},
  \bibinfo{journal}{Nucl. Phys.} \textbf{\bibinfo{volume}{B742}},
  \bibinfo{pages}{208} (\bibinfo{year}{2006}{\natexlab{d}}),
  \eprint{hep-ph/0601165}.

\bibitem[{\citenamefont{Laporta}(2002)}]{Laporta:2002pg}
\bibinfo{author}{\bibfnamefont{S.}~\bibnamefont{Laporta}},
  \bibinfo{journal}{Phys. Lett.} \textbf{\bibinfo{volume}{B549}},
  \bibinfo{pages}{115} (\bibinfo{year}{2002}), \eprint{hep-ph/0210336}.

\bibitem[{\citenamefont{Fleischer and Tarasov}(1994)}]{Fleischer:1994ef}
\bibinfo{author}{\bibfnamefont{J.}~\bibnamefont{Fleischer}} \bibnamefont{and}
  \bibinfo{author}{\bibfnamefont{O.~V.} \bibnamefont{Tarasov}},
  \bibinfo{journal}{Z. Phys.} \textbf{\bibinfo{volume}{C64}},
  \bibinfo{pages}{413} (\bibinfo{year}{1994}), \eprint{hep-ph/9403230}.

\bibitem[{\citenamefont{Baikov and Broadhurst}(1995)}]{Baikov:1995ui}
\bibinfo{author}{\bibfnamefont{P.~A.} \bibnamefont{Baikov}} \bibnamefont{and}
  \bibinfo{author}{\bibfnamefont{D.~J.} \bibnamefont{Broadhurst}}
  (\bibinfo{year}{1995}), \eprint{hep-ph/9504398}.

\bibitem[{\citenamefont{Faisst et~al.}(2004{\natexlab{b}})\citenamefont{Faisst,
  Chetyrkin, and K{{\"u}}hn}}]{ChFK:LL04}
\bibinfo{author}{\bibfnamefont{M.}~\bibnamefont{Faisst}},
  \bibinfo{author}{\bibfnamefont{K.~G.} \bibnamefont{Chetyrkin}},
  \bibnamefont{and} \bibinfo{author}{\bibfnamefont{J.~H.}
  \bibnamefont{K{{\"u}}hn}}, \bibinfo{journal}{Nucl. Phys. Proc. Suppl.}
  \textbf{\bibinfo{volume}{135}}, \bibinfo{pages}{307}
  (\bibinfo{year}{2004}{\natexlab{b}}).

\bibitem[{\citenamefont{Faisst}(2005)}]{Faisst:diss}
\bibinfo{author}{\bibfnamefont{M.}~\bibnamefont{Faisst}}, \bibinfo{journal}{PhD
  thesis, Cuvillier Verlag, Goettingen} \textbf{\bibinfo{volume}{ISBN
  3-86537-506-5}} (\bibinfo{year}{2005}).

\bibitem[{\citenamefont{Gorishnii et~al.}(1989)\citenamefont{Gorishnii, Larin,
  Surguladze, and Tkachov}}]{Gorishnii:1989gt}
\bibinfo{author}{\bibfnamefont{S.~G.} \bibnamefont{Gorishnii}},
  \bibinfo{author}{\bibfnamefont{S.~A.} \bibnamefont{Larin}},
  \bibinfo{author}{\bibfnamefont{L.~R.} \bibnamefont{Surguladze}},
  \bibnamefont{and} \bibinfo{author}{\bibfnamefont{F.~V.}
  \bibnamefont{Tkachov}}, \bibinfo{journal}{Comput. Phys. Commun.}
  \textbf{\bibinfo{volume}{55}}, \bibinfo{pages}{381} (\bibinfo{year}{1989}).

\bibitem[{\citenamefont{Steinhauser}(2001)}]{Steinhauser:2000ry}
\bibinfo{author}{\bibfnamefont{M.}~\bibnamefont{Steinhauser}},
  \bibinfo{journal}{Comput. Phys. Commun.} \textbf{\bibinfo{volume}{134}},
  \bibinfo{pages}{335} (\bibinfo{year}{2001}), \eprint{hep-ph/0009029}.

\bibitem[{\citenamefont{Harlander et~al.}(1998)\citenamefont{Harlander,
  Seidensticker, and Steinhauser}}]{Harlander:1997zb}
\bibinfo{author}{\bibfnamefont{R.}~\bibnamefont{Harlander}},
  \bibinfo{author}{\bibfnamefont{T.}~\bibnamefont{Seidensticker}},
  \bibnamefont{and}
  \bibinfo{author}{\bibfnamefont{M.}~\bibnamefont{Steinhauser}},
  \bibinfo{journal}{Phys. Lett.} \textbf{\bibinfo{volume}{B426}},
  \bibinfo{pages}{125} (\bibinfo{year}{1998}), \eprint{hep-ph/9712228}.

\bibitem[{\citenamefont{Seidensticker}(1999)}]{Seidensticker:1999bb}
\bibinfo{author}{\bibfnamefont{T.}~\bibnamefont{Seidensticker}}
  (\bibinfo{year}{1999}), \eprint{hep-ph/9905298}.

\bibitem[{\citenamefont{Boughezal and Czakon}(2006)}]{Boughezal:2006xk}
\bibinfo{author}{\bibfnamefont{R.}~\bibnamefont{Boughezal}} \bibnamefont{and}
  \bibinfo{author}{\bibfnamefont{M.}~\bibnamefont{Czakon}}
  (\bibinfo{year}{2006}), \eprint{hep-ph/0606232}.

\bibitem[{\citenamefont{Kataev and Kumano}(2003)}]{Kataev:2002sj}
\bibinfo{author}{\bibfnamefont{A.~L.} \bibnamefont{Kataev}} \bibnamefont{and}
  \bibinfo{author}{\bibfnamefont{S.}~\bibnamefont{Kumano}},
  \bibinfo{journal}{J. Phys.} \textbf{\bibinfo{volume}{G29}},
  \bibinfo{pages}{1925} (\bibinfo{year}{2003}), \eprint{hep-ph/0211052}.

\bibitem[{\citenamefont{Kulagin}(2003)}]{Kulagin:2003wz}
\bibinfo{author}{\bibfnamefont{S.~A.} \bibnamefont{Kulagin}},
  \bibinfo{journal}{Phys. Rev.} \textbf{\bibinfo{volume}{D67}},
  \bibinfo{pages}{091301} (\bibinfo{year}{2003}), \eprint{hep-ph/0301045}.

\bibitem[{\citenamefont{McFarland and Moch}(2003)}]{McFarland:2003jw}
\bibinfo{author}{\bibfnamefont{K.~S.} \bibnamefont{McFarland}}
  \bibnamefont{and} \bibinfo{author}{\bibfnamefont{S.-O.} \bibnamefont{Moch}}
  (\bibinfo{year}{2003}), \eprint{hep-ph/0306052}.

\bibitem[{\citenamefont{Diener et~al.}(2005)\citenamefont{Diener, Dittmaier,
  and Hollik}}]{Diener:2005me}
\bibinfo{author}{\bibfnamefont{K.~P.~O.} \bibnamefont{Diener}},
  \bibinfo{author}{\bibfnamefont{S.}~\bibnamefont{Dittmaier}},
  \bibnamefont{and} \bibinfo{author}{\bibfnamefont{W.}~\bibnamefont{Hollik}},
  \bibinfo{journal}{Phys. Rev.} \textbf{\bibinfo{volume}{D72}},
  \bibinfo{pages}{093002} (\bibinfo{year}{2005}), \eprint{hep-ph/0509084}.

\bibitem[{\citenamefont{Erler}(2006)}]{Erler:2006vt}
\bibinfo{author}{\bibfnamefont{J.}~\bibnamefont{Erler}} (\bibinfo{year}{2006}),
  \eprint{hep-ph/0604035}.

\end{thebibliography}
\end{document}